\documentclass[aps,manuscript,showpacs]{revtex4}
\usepackage{graphicx}

\begin{document}

\widetext
{\hfill LBNL-59058}\\

\title{Anti-Lambda polarization in high energy $pp$ collisions with polarized beams}
\author{Qing-hua Xu$^{a,b}$, Zuo-tang Liang$^{a,b}$, and Ernst Sichtermann$^{b}$}
\affiliation{$^a$Department of Physics, Shandong University, Jinan, Shandong 250100, China\\
$^b$Nuclear Science Division, MS 70R0319,
Lawrence Berkeley National Laboratory, Berkeley, CA 94720}


\begin{abstract}

We study the polarization of the anti-Lambda particle in 
polarized high energy $pp$ collisions at large transverse momenta.
The anti-Lambda polarization is found to be sensitive 
to the polarization of the anti-strange sea of the nucleon. 
We make predictions using different parameterizations of 
the polarized quark distribution functions.
The results show that the measurement of longitudinal anti-Lambda polarization 
can distinguish different parameterizations, and  
that similar measurements in the transversely polarized case 
can give some insights into the transversity distribution of 
the anti-strange sea of nucleon.

\end{abstract}


\maketitle


The polarizations of hyperons, in particular the Lambda ($\Lambda$),
have been widely used to study various aspects of spin effects
in high energy reactions for their 
self spin-analyzing parity violating decay\cite{tdlee}.
Many studies, both experimentally 
\cite{ALEPH96,OPAL98,NOMAD00,HERMES,E665,COMPASS} 
and theoretically  \cite{GH93,Burkardt:1993zh,
BL98,Kot98,Florian98,Ma:1998pd,
LL00,XLL02,LL02,XL04,Dong:2005ea,
Ma:1999wp,Ellis2002}
have been made recently,
in particular on the 
spin transfer in high energy fragmentation processes. 
Here, it is of particular interest to know whether 
the SU(6) wave-function or the results drawn from 
polarized deep-inelastic lepton-nucleon scattering (DIS)  
should be used in connecting the spin of the fragmenting 
quark and that of the produced hadrons. 
In addition such studies can give
insight into the spin structure of the nucleon.
For example, it has been pointed out that 
transversity distribution can be studied
by measuring the polarization of the Sigma ($\Sigma^+$) in  
$pp\to\Sigma^+X$ with transversely polarized beams 
and the gluon helicity distributions 
in $pp\to\gamma\Sigma^+X$ with longitudinally polarized beams \cite{XL04}.

Presently, most of our knowledge on the flavor decomposition 
of the proton spin originates from deep-inelastic measurements. 
Polarized inclusive deep-inelastic scattering data from CERN, SLAC, DESY,
and JLAB \cite{DIScern,DISslac,DISdesy,DISjlab}, 
combined with hyperon $\beta$ decay measurements, 
indicate that the strange sea in the nucleon, 
$\Delta s+\Delta\bar s$, is negatively polarized.
Recent semi-inclusive deep-inelastic scattering data \cite{HERMESsemi} 
may indicate a different outcome.
These data do not rely on hyperon decay measurements,
but cover a smaller kinematic range than the 
inclusive data and some analysis aspects 
have come under discussion \cite{Leader}.
Further similar measurements are underway or planned \cite{COMPASSproposal,JiangX}.
Data from elastic neutrino scattering \cite{BNLE734} 
lack the precision to distinguish, 
but better measurements have been proposed \cite{FINESSE}.
Other measurements are called for. 

In this note, we evaluate 
the polarization of inclusive anti-Lambda's in 
polarized $pp$ collisions at large transverse momenta ($p_T$).
We study the dependence of the results on the polarized 
quark distributions and show that the anti-Lambda polarization is quite sensitive to 
the anti-strange sea polarization ($\Delta\bar s$) in the nucleon 
in regions accessible to experiments. 

We consider the inclusive production of anti-Lambda ($\bar\Lambda$) particles with high 
transverse momenta $p_T$ 
in $pp$ collisions with one beam longitudinally polarized.
The $\bar\Lambda$ polarization is defined as,
\begin{equation}
P_{\bar\Lambda}\equiv \frac
{d\sigma{(p_+p \to  \bar\Lambda_+X)}-d\sigma{(p_+p \to  \bar\Lambda_-X)}}
{d\sigma{(p_+p \to  \bar\Lambda_+X)}+d\sigma{(p_+p \to  \bar\Lambda_-X)}}
= \frac {d\Delta \sigma}{d\eta} (\vec pp \to  \bar\Lambda X) /
\frac {d\sigma}{d\eta}(pp \to  \bar\Lambda X),
\label{gener1}
\end{equation}
where $\eta$ is the pseudo-rapidity of the $\bar\Lambda$, and 
$\Delta \sigma$ and $\sigma$ are the polarized and unpolarized 
inclusive production cross sections; 
the subscript $+$ or $-$ denote the helicity of the particle.
We assume that $p_T$ is high enough so that
the factorization theorem is expected to hold and 
the produced $\bar\Lambda$'s are the fragmentation products of high $p_T$ partons 
in $2\to 2$ hard scattering ($ab\to cd$) with one initial parton polarized.
Hence, 
\begin{equation}
\frac {d\Delta \sigma}{d\eta}{(\vec pp \to  \bar\Lambda X)}
=\int_{p_T^{min}}dp_T
\sum_{abcd}\int dx_a dx_b \Delta f_a(x_a)f_b(x_b)
 \Delta D_c^{\bar\Lambda}(z)
D^{\vec ab\to \vec cd}(y) \frac {d \hat {\sigma}}{d\hat t}{( a b\to c d)}
\label{desig2}
\end{equation}
where the sum concerns all possible subprocesses; 
the transverse momenta $p_T$ of the $\bar\Lambda$ is
integrated above $p_T^{min}$;
$\Delta f_a(x_a)$ and $f_b(x_b)$ are the longitudinally
polarized and unpolarized parton distribution functions in the proton 
(whose scale dependence we omit for notational clarity);
$x_a$ and $x_b$ are the momentum fractions carried
by partons $a$ and $b$; 
$D^{\vec ab\to\vec cd}(y)\equiv d\Delta\hat\sigma/d\hat\sigma$ is the 
partonic spin transfer factor in the elementary hard process $\vec ab\to\vec cd$;
$\Delta D_c^{\bar\Lambda}(z)$ is the polarized fragmentation
function defined by, 
\begin{equation}
\Delta D_c^{\bar\Lambda} (z) \equiv
D_{c}^{\bar\Lambda}(z,+)-D_{c}^{\bar\Lambda}(z,-),
\end{equation}
in which the arguments $+$ and $-$
denote that the produced $\bar\Lambda$ has the same or opposite 
helicity as the fragmenting parton $c$.
Experimentally, such spin dependent fragmentation functions can 
be studied in $e^+e^-$-annihilation, polarized deeply inelastic 
scattering and high $p_T$ hadron production in polarized $pp$ 
collisions by measuring hyperon polarization 
in the final states.\cite{GH93,Burkardt:1993zh,
BL98,Kot98,Florian98,Ma:1998pd,LL00,XLL02,LL02,XL04,Dong:2005ea,
Ma:1999wp,Ellis2002}. 
This is because in all these cases, the polarization of quarks 
or anti-quarks before fragmentation can be calculated using 
the standard model for electro-weak interaction or pQCD together 
with the empirical knowledge for polarized 
parton distributions\cite{GRSV2000}.    
The partonic spin transfer factor $D^{\vec ab\to\vec cd}$
is calculable in pQCD and 
turns out to be a function of only $y\equiv p_b\cdot(p_a-p_c)/p_a\cdot p_b$, 
where $p_{a-d}$ are the parton momenta (see e.g. \cite{XLL02}).
The unpolarized cross section $d\sigma/d\eta$
is described by an expression similar to that in Eq.(\ref{desig2})
and can be evaluated from
parameterizations of the unpolarized parton distribution and fragmentation functions.

The unknowns in Eq.(\ref{desig2}) are in principle the polarized 
fragmentation functions $\Delta D_c^{\bar\Lambda}(z)$ and the 
polarized parton distributions $\Delta f_a(x_a)$. 
With external input for one of them, the other 
can be studied via the measurements of $P_{\bar\Lambda}$.

Studies of the polarized fragmentation functions 
to hyperons ($H$) and anti-hyperons ($\bar H$),
$\Delta D^H_c(z)$ and $\Delta D^{\bar H}_c(z)$, 
have been made over the past decade \cite{GH93,Burkardt:1993zh,
BL98,Kot98,Florian98,Ma:1998pd,LL00,XLL02,LL02,XL04,Dong:2005ea,
Ma:1999wp,Ellis2002}.
In particular, the polarized fragmentation functions 
have been calculated \cite{GH93,BL98,LL00,XLL02,LL02,XL04,Dong:2005ea},
for directly produced (anti-)hyperons 
that contain the fragmenting parton $c$ using different models 
for the spin transfer factor $t^F_{H,c}$ ($t^F_{\bar H,c}$)  
from the parton $c$ to the hyperon $H$ (or $\bar H$).
Although data are still too scarce to adequately constrain the models, the 
$z$-dependence of $\Delta D_c^H(z)$ appears to be 
determined by the interplay of the different contributions to 
the unpolarized fragmentation functions,  
as seen from the detailed analysis in Ref.\cite{Dong:2005ea}.
The various models differ mainly by a constant factor.
Besides, a large fraction of high $p_T$ hyperons, 
in particular $\Sigma^+$,  in $pp$ collisions 
are the leading particles in the high $p_T$ jets. 
They have a large probability to be the first rank hadrons in 
the fragmentation of the hard scattered quarks \cite{XLL02}.
This further reduces the influence of the different models 
for $\Delta D_c^H(z)$ on the hyperon polarization.
Measurements of hyperon polarization can thus give insight 
into the polarized parton distributions. 
Examples of this kind have been given in Ref.\cite{XL04}.  

The production of high $p_T$ 
$\Lambda$'s in $pp$ collisions is more involved than $\Sigma^+$ production.
This is because their  
production is dominated by $u$-quark fragmentation, 
and the $u$-quarks contribute at best only a small fraction of the $\Lambda$ spin.
In addition the contribution from decays of heavier hyperons to $\Lambda$'s is sizable.
The resulting $\Lambda$ polarization is expected to be small and its evaluation 
is prone to many uncertainties \cite{XLL02}.

The situation for the $\bar\Lambda$ is different because 
anti-quark fragmentation dominates its production.
The contributions from $\bar u$, $\bar d$ and $\bar s$ 
to the production of jets are expected to be approximately equal.
Since there is a strange suppression factor\cite{Hofmann88} 
of $\lambda\approx 0.3$ for $\bar\Lambda$ production 
in $\bar u$ or $\bar d$ fragmentation compared to $\bar s$, 
we expect that $\bar s$ fragmentation 
gives the most important contribution to $\bar\Lambda$ production 
in $pp\to\bar\Lambda X$. 
In this case, we should expect that many of the 
$\bar\Lambda$'s at high $p_T$ are directly produced 
and contain the hard scattered $\bar s$.

We have made estimates using the Monte-Carlo event 
generator {\sc pythia}6.205\cite{PYTHIA} in its default tune. 
Fig.1 shows the expected fractional contributions to 
$\bar\Lambda$ with $p_T\ge 8$GeV 
as a function of the pseudo-rapidity $\eta$ 
in $pp$ collisions at $\sqrt{s}=200$GeV.
In Fig. 2, we show the fractional contributions 
in the rapidity region $|\eta|<1$ as a function of $p_T$.
We see that, in particular in the region $p_T\ge 8$GeV and $|\eta|<1$, 
$\bar s$ fragmentation indeed provides 
the largest contribution to the $\bar\Lambda$ production, 
whereas the fragmentation contributions 
from $\bar u$ and $\bar d$ are very small.  
In the polarized case, we take the fact that 
the spin transfer factor from 
$\bar s$ to $\bar\Lambda$ is much larger than that 
from $\bar u$ or $\bar d$ into account and 
expect an even stronger $\bar s$  
dominance in the $\bar\Lambda$ polarization. 
Therefore, we expect that in $\vec pp\to\bar\Lambda X$, 
the polarization of the $\bar\Lambda$ should be sensitive to 
the anti-strange sea polarization and that the size 
should be somewhat larger than that of the $\Lambda$. 

Using different sets of parameterizations
for the polarized parton distributions\cite{GRSV2000}  
and the parameterization for the 
unpolarized parton distributions in Ref. \cite{GRV98},
we evaluated $P_{\bar\Lambda}$ as a function of $\eta$ for 
$p_T\ge 8$GeV using Eqs.(1-3). 
As in \cite{BL98,LL00,XLL02,LL02,XL04,Dong:2005ea}, 
we used the SU(6) and DIS pictures for 
the spin transfer factor $t^F_{\bar H,c}$ from parton $c$ 
to the hyperon $\bar H$ that contains the fragmenting $c$. 
The decay contributions from heavier anti-hyperons 
are taken into account in the same way as 
in \cite{BL98,LL00,XLL02,LL02,XL04,Dong:2005ea}.
The second rank anti-hyperons and those from gluon fragmentation 
are taken as unpolarized.  
The different contributions to the $\bar\Lambda$ production 
are calculated using {\sc pythia}6.205. 
The factorization scale is taken as $p_T$.

After the calculations, we found out that 
$|P_{\bar\Lambda}|$ is indeed somewhat 
larger than $|P_\Lambda|$ 
obtained in \cite{XLL02} using the same sets 
of polarized parton distribution functions.
The difference between the results obtained 
using different spin transfer models is relatively small 
whereas the difference between the results obtained using 
different parameterizations of the polarized parton distributions 
can be quite large. 
The latter difference originates predominantly from 
the differences in the parameterizations for 
$\Delta \bar s$ in the $x$ region $0.05<x<0.25$ 
from which most $\bar\Lambda$'s with $p_T>8$GeV/c originate. 
As examples, we show the results obtained using 
the GRSV ``standard'' and ``valence'' sets of 
parametrization of the polarized parton 
distributions\cite{GRSV2000} in Fig.3. 
The influence from the differences in 
$\Delta\bar u$ or $\Delta\bar d$ is very small since
aforementioned fragmentation and spin contributions are small. 
We have cross-checked that $P_{\bar\Lambda}$ 
evaluated with $\Delta\bar u=\Delta\bar d=0$ 
shows no visible difference from the results in  Fig 3.
In view of the current status of our knowledge on 
$\Delta\bar s(x)$ in nucleon in particular the large 
difference between the different sets of parameterizations\cite{GRSV2000}, 
the measurements of $\bar\Lambda$ polarization should 
provide useful information in 
distinguishing these different parameterizations.
   
In the calculations, we chose $p_T\ge 8$GeV so that  
the factorization theorem and pQCD calculations 
are expected to apply. 
We expect that the qualitative features of the results are 
similar at lower $p_T$.
We used the spin transfer factors for the 
hard elementary processes to the leading order (LO) in pQCD. 
This is to be consistent with the fragmentation functions 
where the empirical knowledge is used. 
Clearly, NLO corrections can and should be studied in particular 
in view of the results for $A_{LL}$ as discussed e.g. 
in \cite{Florian2003}.
Such a study can be performed if we know the polarized 
fragmentation functions to this order. 
The fragmentation functions have to be extracted from experiments 
and the presently available data in this connection is still too scare 
for such a study. 

To see what we expect at RHIC \cite{Bunce:2000uv}, 
we have made an estimation of total number of $\bar\Lambda$ events 
and the uncertainties in extracting its polarization. 
Given an integrated luminosity of 100 pb$^{-1}$ and 
a beam polarization of 50\%, we obtained about 100K 
$\bar\Lambda$'s with $p_T>8$ GeV and 
in the pseudo-rapidity range $-1<\eta<1$.
This is well within the parameteres for RHIC spin program\cite{Bunce:2000uv}, 
and corresponds to a statistical uncertainty of 
about $1\%$ for the asymmetry.
We thus conclude that RHIC should have a good chance 
to distinguish between the predictions based on 
different parameterizations of the polarized quark distributions. 
This provides a tool to study the 
polarization of the nucleon sea.

We can extend the calculations to the transversely polarized case, 
where we have, 
\begin{equation}
\frac {d\delta \sigma}{d\eta}{(p^\uparrow p \to  \bar\Lambda X)}
=\int_{p_T^{min}}dp_T
\sum_{abcd}\int dx_a dx_b \delta f_a(x_a)f_b(x_b)
 \delta D_c^{\bar\Lambda}(z)
D_T^{a^\uparrow b\to c^\uparrow d}(y) 
\frac {d \hat {\sigma}}{d\hat t}{( a b\to c d)}
\label{desig2tr}
\end{equation}
Here, $\delta D_c^H(z)$ and $\delta q(x)$ 
are the polarized fragmentation functions in the transversely polarized case 
and transversity distributions of the quarks or anti-quarks.
They can be studied experimentally e.g. in 
semi-inclusive deep-inelastic lepton-nucleon 
scattering with transversely polarized nucleon 
or high-$p_T$ hyperon production in 
transversely polarized $pp$ collisions 
by measuring the hyperon polarization in the final state.   
The partonic spin transfer factor for 
the elementary hard scattering process 
is replaced by $D_T^{a^\uparrow b\to c^\uparrow d}(y)$ for transverse polarization, which 
is also calculable from pQCD for the elementary hard scattering  
processes (see e.g. \cite{Collins:1993kq}).

As we have emphasized before, 
the results shown in Figs.1 and 2 are for unpolarized reactions. 
Just as in the longitudinally polarized case, 
the $\bar\Lambda$ polarization in the transversely polarized case 
is also dominated by the $\bar s$ fragmentation and spin contributions,
and therefore $P_{\bar\Lambda,T}$ should be sensitive to $\delta\bar s(x)$. 
We made an estimate of $P_{\bar\Lambda,T}$ assuming 
$\delta D_c^H(z)=\Delta D_c^H(z)$ and 
$\delta q(x)=\Delta q(x)$. 
The results are given in Fig.4. 
The differences between the results in 
Fig.3 and 4 originate from the differences between 
$D^{\vec ab\to \vec cd}(y)$ and 
$D_T^{a^\uparrow b\to c^\uparrow d}(y)$.

In summary, 
we studied anti-Lambda ($\bar{\Lambda}$) polarization ($P_{\bar\Lambda}$) 
in polarized high energy $pp$ collisions at high transverse momenta $p_T$. 
A large part of the centrally produced $\bar\Lambda$'s at high $p_T$ 
are found to originate from anti-strange quark ($\bar s$) fragmentation.
Therefore, the anti-Lambda polarization $P_{\bar\Lambda}$ is sensitive to the 
polarization of the anti-strange sea in nucleon. 
Measurements of $P_{\bar\Lambda}$ 
in longitudinally polarized $pp$ collisions can provide  
new insights in the spin-dependent quark distributions in 
particular $\Delta \bar s(x)$. 
In the transversely polarized case, similar studies can give some insights 
into the transversity distribution $\delta\bar s(x)$.

The studies can be extended to other (anti-)hyperons in a straightforward way. 
If high accuracy  measurements can be carried out, 
this may provide a complementary path to flavor decomposition of the nucleon spin.
  
\vspace{0.3cm}
{\bf Acknowledgments}

This work was supported in part by the United States Department of Energy and by 
the National Science Foundation of China (NSFC) with grant No. 10175037 and
No. 10405016.

\widetext

\begin{figure}[htb!]
\resizebox{6.0in}{4.0in}{\includegraphics{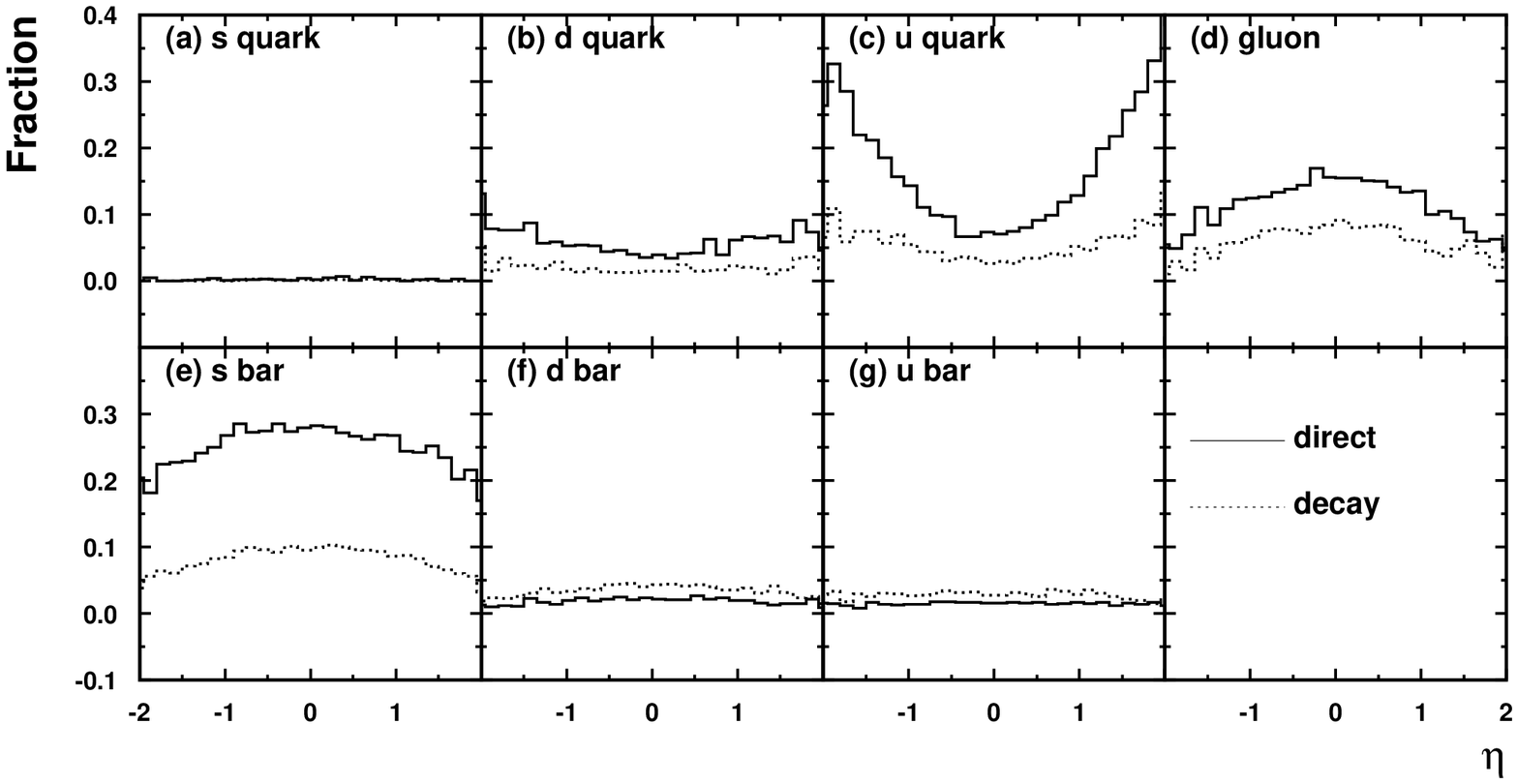}}
\caption{
Contributions to $\bar\Lambda$ production at $p_T\ge 8$ GeV/c 
in $pp$ collisions at $\sqrt s=200$ GeV.
The solid and dashed lines are respectively 
the directly produced and decay contributions.  
}
\label{fig1}
\end{figure}

\begin{figure}[htb!]
\resizebox{6.0in}{4.8in}{\includegraphics{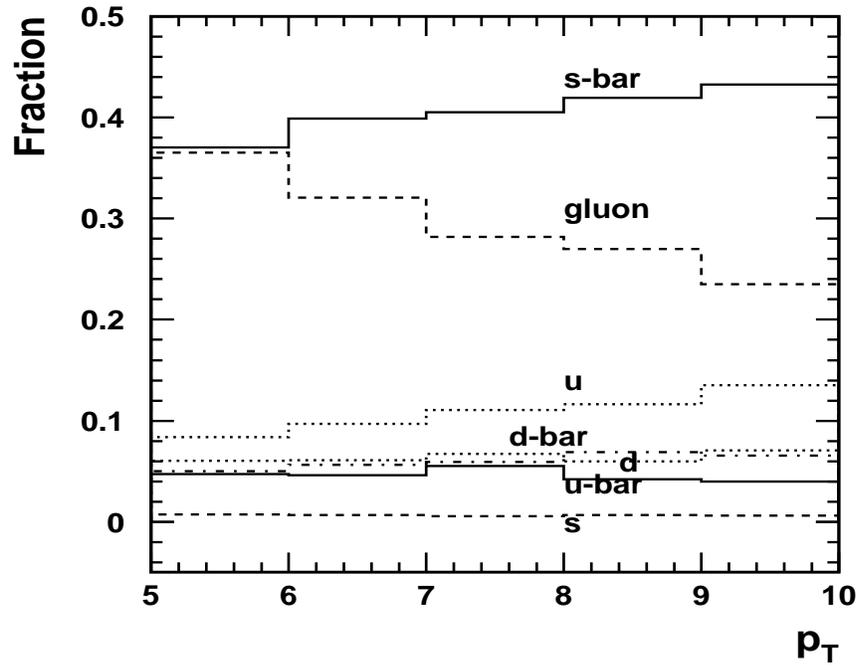}}
\caption{
Contributions to $\bar\Lambda$ production for 
$|\eta|<1$ in $pp$ collisions at $\sqrt s=200$ GeV versus transverse momentum $p_T$.}
\label{fig2}
\end{figure}

\begin{figure}
\resizebox{6.0in}{5.0in}{\includegraphics{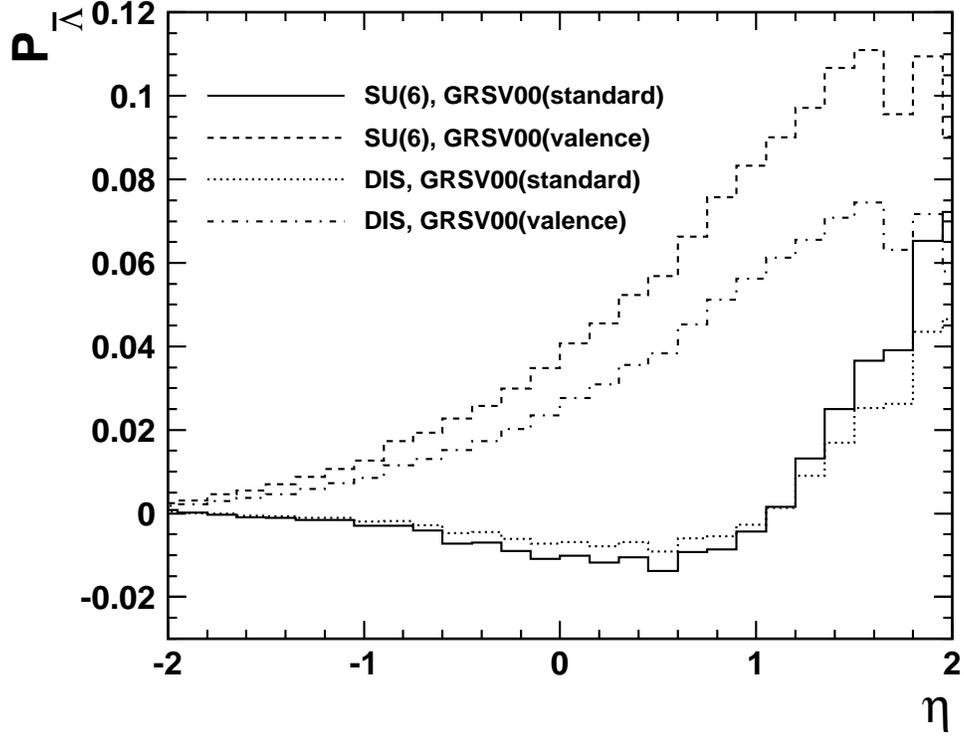}}
\caption{ 
Longitudinal $\bar\Lambda$ polarization for transverse momentum 
$p_T\ge 8$ GeV/c in $pp$ collisions at $\sqrt s=200$ 
with one longitudinally polarized beam versus pseudo-rapidity 
$\eta$ of the $\bar\Lambda$.
Positive $\eta$ is taken along the polarized beam direction.  
}
\label{fig3}
\end{figure} 

\begin{figure}
\resizebox{6.0in}{5.0in}{\includegraphics{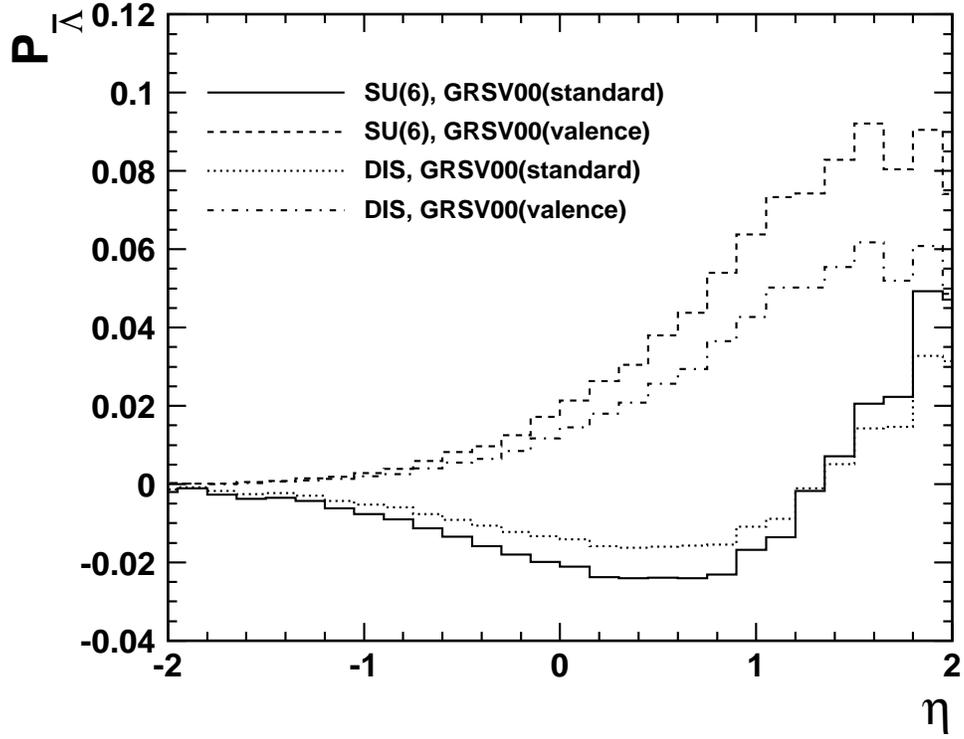}}
\caption{ 
Transverse $\bar\Lambda$ polarization for transverse momentum 
$p_T\ge 8$ GeV/c in $pp$ collisions at $\sqrt s=200$ 
with one transversely polarized beam versus pseudo-rapidity 
$\eta$ of the $\bar\Lambda$.
Positive $\eta$ is taken along the polarized beam direction.  
}
\label{fig4}
\end{figure} 

\end{document}